\documentclass[pre,twocolumn,showpacs]{revtex4}
\usepackage{amssymb}
\usepackage{graphicx}
\usepackage{amsmath}

\begin{document}


\input{epsf}
\title{A density functional study of pressure induced superconductivity in P and its
implication for spintronics.}
\author{S.~Ostanin$^1$, V.~Trubitsin$^2$, 
 J.B.~Staunton$^1$, S.Y.~Savrasov$^3$}
\address{$^1$Department of Physics, University of 
 Warwick, Coventry CV4 7AL, United Kingdom}
\address {$^2$Physico-Technical Institute,
 Ural Branch of RAS, 132 Kirov Str., 426001 Izhevsk, Russia}
\address {$^3$Department of Physics, New Jersey Institute of Technology,
 Newark, New Jersey 07102}

\date{\today}

\begin{abstract}
 The stability of high-pressure phases of P has been studied using density
 functional theory and the local density approximation. 
 Using a linear response technique, we have calculated the 
 phonon spectrum and electron-phonon interaction for bcc P 
 and predict it to be superconducting
 with $T_c$ of 19 K. We propose that this phase might be realized
 in epitaxial thin films using templates such as 
 V(100), Fe(100) or Cr(100) relevant to spintronics applications.
\end{abstract}

\pacs{61.50.Ks, 74.25.Jb, 74.78.Fk, 85.75.-d}

\maketitle



 The study of properties of materials under pressure is a key experimental 
 technique and recent discoveries of superconductivity in such simple 
 metals as compressed Fe \cite{NatureFe} and Li \cite{NatureLi} are 
 currently generating enormous interest \cite{NatureLi1,FeWorks,FeWorks1}. 
 Black insulating phosphorus, with an orthorhombic A17 structure, which is the 
 most stable form at ambient conditions, exhibits a series of pressure 
 induced phase transitions. \cite{Kikegawa,Akahama,Akahama2,Kawamura} The first
 is a transformation to a metallic phase with the simple cubic structure at 
 10 GPa. This phase is known to superconduct with $T_c$=10K \cite{Kawamura}. 
 An observation of a new simple bcc structure above 262 GPa 
 has been made recently \cite{Akahama}, which points out that similar or even 
 stronger electron-phonon interaction and pressure induced superconductivity 
 may be also realized in bcc P.
 Growth of P containing materials by thin film techniques has been developed 
 very recently \cite{Pearton} to fabricate the class of wide gap phosphide 
 ferromagnetic semiconductors which may provide an ideal opportunity to
 stabilize the bcc phase of P at ambient conditions using suitable templates 
 such as the V(100), Fe(100) or Cr(100) substrates. This may lead to a 
 significant breakthrough in technological spintronics applications in which 
 combined spin and superconducting degrees of freedom will provide a 
 new level of functionality for microelectronic devices. 

 In this work we address both the issue of the phase stability of P as well as 
 the superconductivity of its highly pressurized bcc structure. 
 We use density functional total energy \cite{DFTReview} and linear response 
 \cite{BaroniReview} techniques which have proven to provide a reliable
 description of various ground state properites for a large class of materials. 
 We also study the electronic structure of the Fe$_5$/P$_n$/Fe$_4$ ($n$=3,5) 
 superlattices, with the lattice constant of bulk Fe to investigate the problem 
 of how to stabilize the bcc phase under ambient pressure conditions.

 In the past, structural phase transformations in P have been the subject of 
 many experimental \cite{Kikegawa,Akahama,Akahama2,Kawamura}
 and theoretical \cite{Chang,Sasaki,Christensen} density functional theory (DFT) based studies.
 Orthorhombic A17 structured P transforms to the rhombohedral 
 A7 phase at 4.5 GPa.\cite{Kikegawa} 
 The puckered P layers in the ambient-pressure A17 structure have strong covalent 
 intralayer bonding and weak bonding between
 layers that result in an anisotropic compressibility. 
 Under pressure, the distance between the interlayer P atoms in the A17 structure 
 decreases faster than the intraplane distance forming the semimetallic A7 structure.  
 This A7 phase, which is similar to the structure of such group-V elements 
 as As, Sb, and Bi, undergoes a transition to the 
 simple cubic (sc) metallic phase at 10 GPa.\cite{Kikegawa}  
 In the group-V elements, this A7$\to$sc
 transition is observed also in Sb at 7 GPa. 

 With increasing pressure, at 137 GPa, the sc structure transforms 
 into a simple hexagonal (sh) one.\cite{Akahama} 
 This sc$\to$sh transition, which is accompanied by a large volume 
 reduction of 7.6 $\%$, is the first such
 observation reported for an elemental system. The sc-sh transformation
 probably occurs via an intermediate phase \cite{Akahama}, and results 
 in the change in the co-ordination number from 6 to 8.  
 Since the sh structure has a low atomic packing fraction,
 namely 0.605, a phase with a higher packing fraction such as bcc, 
 fcc or hcp should be expected at higher pressures.
 Indeed, with increasing pressure, the $c$/$a$ ratio of the sh phase 
 increases from 0.948 to 0.956, and, finally, above 262 GPa,
 a new high-pressure phase is observed, which has been proposed to be 
 bcc.\cite{Akahama} A high-pressure bcc structure has also been found 
 in Sb at 28 GPa and in Bi at 7.7 GPa.\cite{Aoki} In contrast,  
 the sh phase of Si and Ge transforms to the hcp structure.\cite{Vohra}  
 
 To date, despite these data, no full theoretical description of the phase 
 stability in P under extreme high pressure has been reported.
 There are a few DFT calculations on the
 structural stability of P under pressure.\cite{Chang,Sasaki}
 Using the pseudopotential method, Chang and Cohen \cite{Chang}
 reproduced the A17$\to$A7$\to$sc phase transitions, with
 the crystal energy of the A7 phase being shifted by 2.3 mRy,
 but failed to find any stable 
 closed-packed P phase at high pressure. In contrast, Sasaki {\it et al.}
 \cite{Sasaki} inferred a sc-bcc transition
 at a pressure of 135 GPa, rather far from the recent 
 experimental observations.\cite{Akahama} 
 This paper shows that ab-initio DFT calculations can successfully describe
 all the structural transitions in the phase diagram in P and, in 
 particular, quantifies the sh-bcc transition, the highest-pressure
 structural transformation observed up to now for an elemental material. 

 The use of the Debye-Gr{\"u}neisen theory within 
 DFT calculations is the simplest way to
 investigate structural phase transitions at finite temperatures. 
 The characteristic Debye temperature
 ${\Theta}_{D}$, calculated in terms of the bulk modulus,
 can be used to constuct the free energy $F(T,V)$ as the
 function of ${\Theta}_{D}$ and $V$.
 The successful applicability of this model,
 in despite of its harmonic approximation for lattice 
 vibrations, has been demonstrated 
 \cite{Ost1} for some elemental materals under high pressure.  
 To construct the phase diagram, the thermodynamical Gibbs
 potentials $G(P,T) = F(V,T) + P V$ are calculated and compared for
 various structures while the isothermal pressure dependence $P(V)$ is 
 obtained by direct differentiation
 $P=-{(\partial F/ \partial V)}_{T}$ for each structure.
 We use the full potential linearized augmented plane waves
 (WIEN) code \cite{Blaha}, which is one of the most accurate
 DFT schemes. Exchange and correlation effects
 are treated using the generalized-gradient approximations 
 \cite{PERDEV} while {\it muffin-tin} radii of 1.7 a.u. are
 fixed for all structures.  
 As Fig.~1 of the phase diagram of P
 shows, the calculations reproduce quantitatively well
 all available experimental data on the pressure induced structural
 transformations in P. In particular, at just above the critical 
 pressure where the sh-bcc transition takes place, at 260 GPa
 the bcc phase is a real possibility and the bcc phase remains 
 stable at still higher pressures.
 The free energy calculation of the bcc phase shows
 a minimum at 3.05~\AA . 
 We predict also that the sh-bcc transition has a positive 
 $dT/dP$ slope at room temperature.
\begin{figure}
\resizebox{0.9\columnwidth}{!}{\includegraphics*{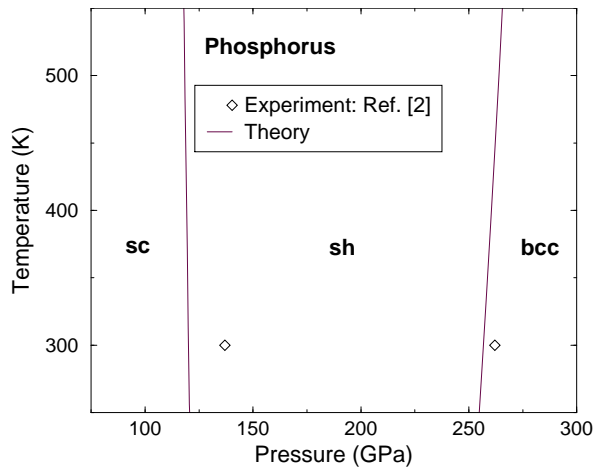}}
\caption{Phase diagram of P at high pressures.}
\label{fig:PHASE_DIAGR}
\end{figure}

 Since the sc phase of P becomes superconducting (SC) below 10~K,
 one may expect similar electron-phonon interaction (EPI) strengths in P
 at higher pressures. Hence, a conventional $s$-wave pairing state may also
 arise in bcc P. We investigate this possibility by using a
 linear-response full-potential LMTO
 DFT method \cite{Savrasov}, which can describe the SC
 properties for $sp$ materials with high accuracy \cite{Kong}
 as well as those of $d$-metals under high pressure.\cite{OTS+}
 The calculated FP-LMTO electronic structure of bcc P is practically 
 identical to that of the WIEN-code calculation.    
 The phonon dispersions ${\omega}_m$($\bf q$) and density of states 
 $F$($\omega$) are obtained from this linear-response technique 
 \cite{Savrasov} and used to calculate the Eliashberg spectral function
 ${\alpha}^2$($\omega$)$F$($\omega$) which defines the EPI strength
 $\lambda$ = ${\int}_0^{\omega _{max}} {\omega}^{-1} \, 
  {\alpha}^2(\omega) \, F(\omega) \, d \omega$. 
 The function ${\alpha}^2$($\omega$)$F$($\omega$), shown in Fig.~2,
 is dominated by the two optical peaks at 2.1 THz and 8.3 THz.
\begin{figure}
\resizebox{0.9\columnwidth}{!}{\includegraphics*{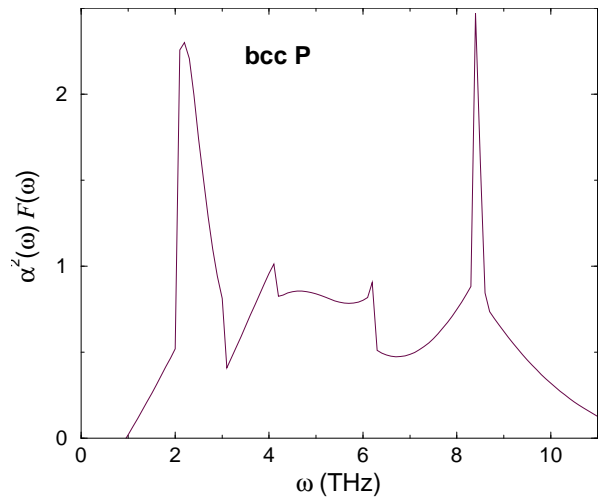}}
\caption{The Eliashberg spectral function of bcc P. }
\label{fig:ELIASHBERG}
\end{figure}  

 The SC transition temperature is calculated by means of the 
 Allen-Dynes modified McMillan expression \cite{Allen} 
 $T_c = \frac {< \omega >}{1.2} \, exp {\left [
 \frac {-1.04 {( 1 + {\lambda}) } }{ {\lambda} -
 { {\mu}^{*} }  {( 1 + 0.62 {\lambda}) } } \right] }$ 
 using the calculated EPI strength and assuming a value
 of the Coulomb pseudopotential of $\mu ^*$=0.15 which was also used
 to calculate $T_c$ for the sc phase of P.  
 In bcc P we find $T_c$=19~K.
 Regarding the transport properties of bcc P in 
 the normal state, the temperature dependence of the specific dc 
 resistivity shows a crossover from power law to linear dependence
 at 50~K.

 There is considerable current interest in technological spintronics
 applications, in which the spin of carriers is exploited
 to provide new functionality for microelectronic devices.
 Growth of P containing materials by thin film techniques, such as
 molecular beam epitaxy or pulsed laser deposition, has been
 developed to fabricate the class of wide gap phosphide ferromagnetic
 semiconductors.\cite{Pearton} 
 Epitaxial thin film structures offer ideal opportunities
 to stabilize the bcc phase of P at ambient conditions using
 suitable growth templates such as the V(100), Fe(100) or Cr(100) substrates.
 Consequently, we now turn to study the electronic and magnetic properties of 
 thin films of P (3-5 layers in thickness) embedded in ferromagnetic (FM)
 bcc-Fe. During the last decade considerable theoretical and experimental
 progress has been made towards the understanding of metallic magnetism   
 in superlattices and multilayers. However, to date, no such investigations
 have been made for Fe/P. 

 The electronic structure of the Fe$_5$/P$_n$/Fe$_4$ ($n$=3,5) 
 superlattices, with the lattice constant of bulk Fe, is calculated 
 from the WIEN-code. Fig.~3 shows the total and layer-resolved electronic
 density of states (DOS) for $n=3$. Evidently the DOS of phosphorus is fairly
 insensitive to monolayer (ML) position and p-d hybridization has broadened
 d-related peaks in the Fe interface DOS. Following on from this 
 Fig.~4 shows the layer-resolved spin magnetic moments and the spin magnetic 
 moment per atom on the Fe 
 interface layer is about 20~$\%$ smaller than that of bulk bcc Fe
 whereas the second Fe layer from the ideally sharp Fe/P interface has 
 an enhanced local spin moment compared to that of bulk Fe.
 The P atoms away from the interface have no induced magnetic moments.  
 The spin magnetic moment on the P interface atoms
 are tiny and antiparallel to the Fe magnetization direction.
\begin{figure}
\resizebox{0.9\columnwidth}{!}{\includegraphics*{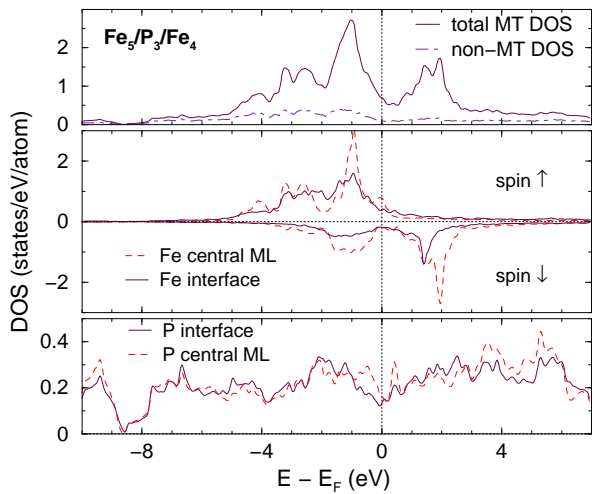}}
\caption{Total and layer-resolved electronic DOS of the 
 Fe$_5$/P$_3$/Fe$_4$ superlattices
 with the lattice constant of bulk Fe.}
\label{fig:DOS}
\end{figure}
\begin{figure}
\resizebox{0.9\columnwidth}{!}{\includegraphics*{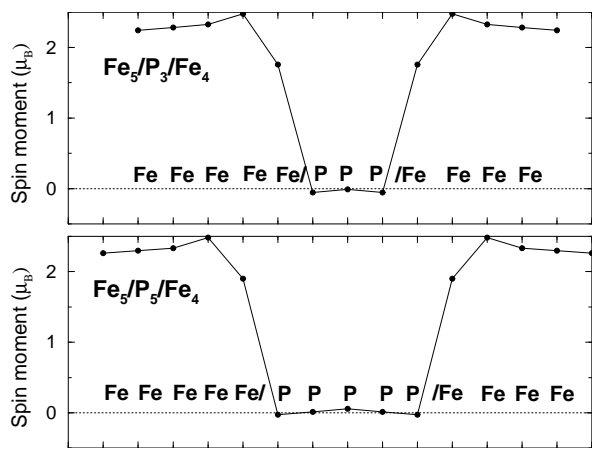}}
\caption{Calculated layer-resolved spin moments of 
 Fe and P in the Fe$_5$/P$_n$/Fe$_4$ ($n$=3,5) superlattices
 with the lattice constant of bulk Fe.}
\label{fig:MOMENT}
\end{figure}
 Thus, the total magnetic moments of the Fe/P/Fe superlattices,
 which are almost independent of the thickness of P, 
 are slightly reduced from those of a relevant pure Fe system 
 due to the AF Fe-P interface coupling. Obviously, this effect would be
 important for superlattices with thin Fe slabs. 
 We note that a similar but more pronounced AF coupling
 takes place at the Fe/V (001) interface.\cite{Hjorvarsson,OstUzd}
 
 The proximity effect in FM/SC heterostructures
 means that the Cooper pair amplitude in an exchange field does not
 decay exponentially to zero on the FM side  
 but oscillates with decreasing amplitude as a function of the
 distance from the interface. These oscillations occurring in what is
 known as the Fulde-Ferrell-Larkin-Ovchinnikov state \cite{FFLO} may lead 
 striking changes in the SC transition temperature (as
 observed in Fe/V and Fe/Nb heterostructures \cite{Wong})
 as a function of the FM slab thickness $d_{Fe}$.
 We suggest that Fe/P heterostructures may show similar effects.

 Another interesting application of the SC state 
 in the stabilized bcc phase of P might be available with V/Fe/P 
 junctions, in which thin FM films are sandwiched between 
 two superconductors with differing SC transition temperatures. 
 Such systems support spontaneous currents parallel to 
 the FM/SC interface while 
 the spin polarization of these currents depends on band filling
 \cite{Krawiec} and, hence, can be readily adjusted. 

 In {\it summary}, our study of P under high pressure
 based upon DFT and Debye model calculations produces 
 a pressure-temperature phase diagram  
 in good agreement with experiment. Pressure-induced
 metallization, attributed to the strong P-P interactions, is
 enhanced in the bcc phase which is stable for pressures 
 in excess of 265 GPa. 
 We have demonstrated that bcc-P might be a $s$-wave 
 superconductor below 19~K owing primarily to strong EPI for 
 the phonon bands along [110]. 
 Finally, we predict that the SC properties of bcc-P can be
 realized in Fe/P heterostructures in which
 the bcc phase of P should be stable.  

 The authors acknowledge support from the EPSRC (U.K.). 
 S.Y.S acknowledges the support from the grants NSF DMR 0238188 
 and NJSGC 02-42.

\end{document}